\documentclass[twocolumn,prl,superscriptaddress]{revtex4}
\usepackage[colorlinks,linkcolor=blue,anchorcolor=blue,citecolor=blue,filecolor=blue,menucolor=blue,runcolor=blue,urlcolor=blue,frenchlinks=blue]{hyperref}
\usepackage{amsmath}
\usepackage{graphicx}
\usepackage{amssymb}
\usepackage{dcolumn}
\usepackage{mathrsfs}
\usepackage{bm}
\usepackage{color}
\begin{document}

\title{New-type geometric gates in atomic arrays without Rydberg blockade}

\author{Yue Ming}
\affiliation {Key Laboratory of Atomic and Subatomic Structure and Quantum Control (Ministry of Education), Guangdong Basic Research Center of Excellence for Structure and Fundamental Interactions of Matter, School of Physics, South China Normal University, Guangzhou 510006, China}

\author{Zhao-Xin Fu}
\affiliation {Key Laboratory of Atomic and Subatomic Structure and Quantum Control (Ministry of Education), Guangdong Basic Research Center of Excellence for Structure and Fundamental Interactions of Matter, School of Physics, South China Normal University, Guangzhou 510006, China}

\author{Yan-Xiong Du}
\email{yanxiongdu@m.scnu.edu.cn}
\affiliation {Key Laboratory of Atomic and Subatomic Structure and Quantum Control (Ministry of Education), Guangdong Basic Research Center of Excellence for Structure and Fundamental Interactions of Matter, School of Physics, South China Normal University, Guangzhou 510006, China}

\affiliation {Guangdong Provincial Key Laboratory of Quantum Engineering and Quantum Materials, Guangdong-Hong Kong Joint Laboratory of Quantum Matter, Frontier Research Institute for Physics, South China Normal University, Guangzhou 510006, China}


\begin{abstract}
The Rydberg blockade effect plays an important role in realizing two-qubit gates in atomic arrays. Meanwhile, such mechanics will increase the crosstalk between atoms and enhance the decoherence. In this paper, we propose a new scheme to realize the controlled-phase gate without Rydberg blockade. The scheme works effectively with large atomic spacings and is insensitive to the thermal motions of atoms. The proposal is robust against random noises due to the geometric characteristic and operates fast based on the non-adiabatic evolution. The proposed gate is actually a new-type geometric gate that consolidates the non-adiabatic holonomic control and the unconventional geometric control simultaneously. The interference between two different types of geometric phases can be investigated. Furthermore, we show that the scheme with weak Rydberg interaction requires much less physical resources than the present Rydberg blockade scheme. Therefore, our proposal provides a fast and robust way to realize geometric quantum control, and it may trigger the discoveries of new geometric gates in high-dimensional Hilbert space.
\end{abstract}

 \maketitle

\textit{Introduction.}--- Trapped atoms in tweezers arrays are attractive physical platforms for large-scale quantum computation and quantum simulations \cite{Saffman2016,Henriet2020,Browaeys2020,Scholl2021,Ebadi2021,Kim2024}. High-fidelity quantum gates that exceed the quantum error correction threshold have also been realized \cite{Levine2019,Madjarov2020,Fu2022,Evered2023,Bluvstein2024}. The Rydberg interactions are of essential importance in the entangling gates on the neutral-atoms platform \cite{Saffman2010,Jaksch2000,Isenhower2010,Jau2016,Shi2018,Liu2020,Theis2016,Wilk2010,Sun2020,Zeng2017,Li2022}. In most reported experiments, the Rydberg blockade effect is adopted, which is operated at a distance of several micrometers. Such a short distance will have a considerable impact on the crosstalk between neighboring qubits. Increasing the distance between the qubits while maintaining the Rydberg blockade condition may reduce crosstalk; nevertheless, higher principal numbers of Rydberg states $n_r$ will be required. This makes the scheme being more sensitive to electromagnetic noise \cite{Evered2023}. Therefore, seeking entangling gates without Rydberg blockade condition and accommodating robust, faster quantum control will be a near-term task for quantum computation with neutral atoms.

On the other hand, geometric phases that are built-in fault-tolerant are frequently adopted in the design of quantum gates \cite{Sjoqvist2008}. For example, Abelian geometric phases can be accumulated during the adiabatic evolution along non-degenerate eigenstates \cite{Berry1984,Moller2008}, while non-Abelian geometric (holonomic) phases can be obtained in degenerate cases \cite{wilzeck1984,duan2001}. The investigations of geometric phases have also been generalized to the non-adiabatic situation \cite{Aharonov1987,SLZhu2002}. Unconventional geometric quantum gates that use the geometric total phase are proposed \cite{zhusl2003}, where the cancellation of the dynamical phase is removed. Non-adiabatic holonomic quantum computation is also proposed by manipulating a three-level system with vanishing dynamical phases \cite{Sjoqvist2012,Feng2013,Abdumalikov2013,Liang2014,Sjoqvist2016,Ma2023}. One may ask whether new types of geometric quantum gates can be further established. Thanks to the flexible controllability of the interacting Hamiltonian in atomic arrays, quantum states inside high-dimensional Hilbert space can be manipulated. Thus geometric quantum gates with unique dynamics can be investigated.

In this Letter, we propose a new protocol for the realization of controlled-phase gates in atomic arrays via Rydberg interactions. Unlike typical gates that work in the Rydberg blockade region, our proposal removes this restriction and thereby realizes two-qubit gates with large atomic spacing and small $n_r$. It is found that the proposed gate consolidates the non-adiabatic holonomic control and the unconventional geometric control upon the corresponding subspace simultaneously. The geometric character validates the robustness against the random noise of the control waveforms. The composite geometric gate is also robust against thermal motion owing to the large separation distance between the atoms. By preparing the system to the superpositions of two computational bases, interference results between two different kinds of geometric phases can be obtained. It is also shown that the scheme with weak Rydberg interaction requires much less physical resources than the present Rydberg blockade scheme. Therefore, our proposal provides a fast and robust way to realize two-qubit gates in atomic arrays and may deepen the knowledge about the geometric control of quantum states.

\textit{Two-qubit gates without Rydberg blockade.}--- In the following, we introduce how to realize two-qubit gates with Rydberg interaction. As depicted in Fig. 1(a), we consider an interacting model of two atoms (labeled $1, 2$), where each atom has three levels $\{|0\rangle, |1\rangle, |r\rangle\}$, $|0\rangle, |1\rangle$ are the ground states, and $|r\rangle$ is the Rydberg state. The state $|1\rangle$ is coupled to $|r\rangle$ with Rabi frequency $\Omega$, detuning $\Delta$ and phase $\varphi$. The interaction strength between two Rydberg states are given by $V$. Under the single-qubit basis, the interacting Hamiltonian is given by
\begin{eqnarray}
&&H=(\frac{\Omega}{2}e^{i\varphi}|1\rangle_1\langle r|\otimes I_2+I_1\otimes|1\rangle_2\langle r|+\mathrm{H.c.})\\
&&+\Delta(|r\rangle_1\langle r|\otimes I_2+I_1\otimes|r\rangle_2\langle r|)+V|r\rangle_1\langle r|\otimes|r\rangle_2\langle r|, \nonumber
\end{eqnarray}
where $I_1=|0\rangle_1\langle 0|+|1\rangle_1\langle 1|+|r\rangle_1\langle r|$, $I_2=|0\rangle_2\langle 0|+|1\rangle_2\langle 1|+|r\rangle_2\langle r|$, $\hbar=1$. The system governed by Hamiltonian (1) will evolve in a Hilbert space of dimension 9, whereas we only concern the dynamics of the computational basis $\{|00\rangle, |01\rangle, |10\rangle, |11\rangle\}$. It can be checked that states $\{|01\rangle, |0r\rangle\}$, $\{|10\rangle, |r0\rangle\}$ and $\{|11\rangle, |R\rangle, |rr\rangle\}$ form subspace respectively and the corresponding Hamiltonian are given by \cite{Chen2024,Sun2023}
\begin{subequations}
\begin{align}
H_{01}&=(\frac{\Omega}{2}e^{i\varphi}|01\rangle\langle 0r|+\mathrm{H.c.})+\Delta|0r\rangle\langle 0r|,\\
H_{10}&=(\frac{\Omega}{2}e^{i\varphi}|10\rangle\langle r0|+\mathrm{H.c.})+\Delta|r0\rangle\langle r0|,\\\nonumber
H_{11}&=[\frac{\sqrt{2}}{2}\Omega e^{i\varphi}(|11\rangle\langle R|+|R\rangle\langle rr|)+\mathrm{H.c.}]\\
&+\Delta|R\rangle\langle R|+(V+2\Delta)|rr\rangle\langle rr|,
\end{align}
\end{subequations}
where $|R\rangle=(|1r\rangle+|r1\rangle)/\sqrt{2}$. The state $|00\rangle$ is decoupled from the other states.

As depicted in Fig. 1(b), an initial state $|\psi_\mathrm{i}\rangle=\sum_ka_k|k\rangle, k=00, 01, 10, 11,$ will evolve along the subspace governed by $H_{01}, H_{10}, H_{11}$ and arrive at the final state $|\psi_\mathrm{f}\rangle$. To realize the controlled-phase gates, a direct strategy is to drive each computational basis cyclically with additional phases accumulated at the end of the evolution. The evolution operator will be derived as
\begin{equation}
    U_{\mathrm{cp}}=\sum_ke^{-i\varphi_k}|k\rangle\langle k|,
\end{equation}
where $\varphi_k$ are the phases accumulated upon the corresponding states after the evolution ($\varphi_{00}=0$). The controlled-phase is defined as $\delta\gamma=\varphi_{11}-\varphi_{10}-\varphi_{01}$ and $\delta\gamma=\pm\pi$ corresponds to the controlled-$Z$ ($\mathrm{C_z}$) gate.

\begin{figure}[ptb]
\begin{center}
\includegraphics[width=8.5cm]{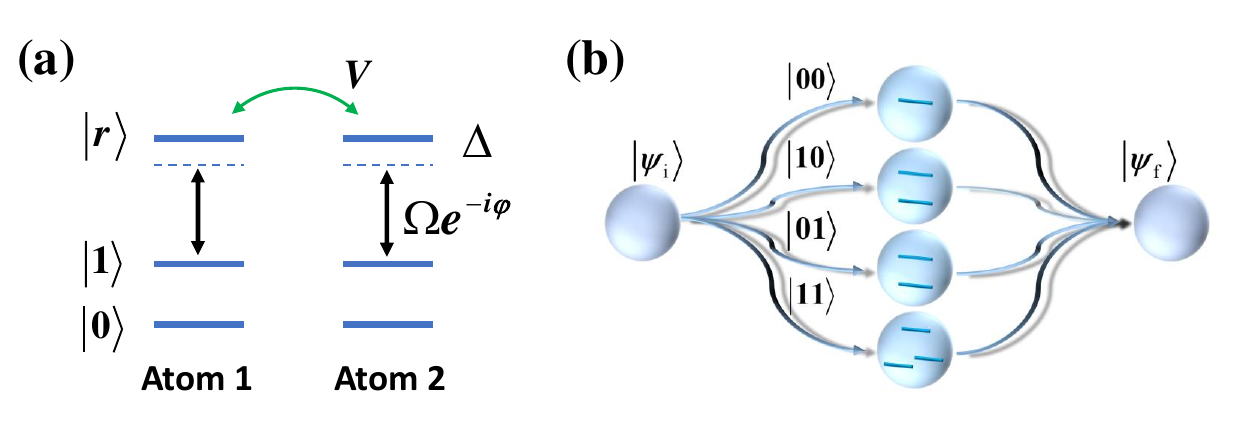}
\caption{\label{fig1}
Scheme of controlled-phase gates based on geometric quantum control. (a) Coupling configuration of two interacting atoms (1 and 2). Three levels of the two atoms are used, where $|0\rangle, |1\rangle$ are the ground states and $|r\rangle$ are the Rydberg states. Both $|1\rangle$ are coupled to $|r\rangle$ with Rabi frequencies $\Omega$, phase $\varphi$, and detuning $\Delta$. The interaction strength between two atoms is given by $V$ which do not need to fulfill the Rydberg blockade condition. (b) Illustration of the mechanics of the controlled-phase gate. For the system prepared in state $|\Psi_\mathrm{i}\rangle$, the two-qubit basis $|k\rangle, k=00, 01, 10, 11,$ will evolve cyclically in its subspace and accumulate relative phases. The new feature of the proposed scheme is that state $|10(01)\rangle$ will be governed by the unconventional geometric control in the two-level system, while state $|11\rangle$ will be governed by the non-adiabatic holonomic control in the three-level system. Therefore, the proposed gate is the mixture of unconventional geometric gate and the non-adiabatic holonomic gate.
}
\end{center}
\end{figure}

Usually, the population of $|01\rangle(|10\rangle)$ and $|11\rangle$ cannot return to the initial ones simultaneously due to asynchronous dynamics. Adiabatic methods and the optimization algorithm have been developed to overcome such problems, nevertheless, may face the obstacles of long evolution time or specific parameter settings \cite{Beterov2016,Pelegri2022,Mitra2020}. To achieve the geometric controlled-phase gate, we adopt the following pulse sequences with the diabatic manners \cite{Kang2018,Jandura2022}:
\begin{equation}
\begin{split}
&t: 0\rightarrow4T, \Omega=\kappa V, \Delta=-V/2;\\
&t: 0\rightarrow T, \varphi=0; t: T\rightarrow2T,  \varphi=\pi/2;\\
&t: 2T\rightarrow3T, \varphi=0; t: 3T\rightarrow4T, \varphi=\pi/2.
\end{split}
\end{equation}
Cyclic evolution is found to be achieved within a wide range of parameters that meet the condition $\sqrt{4\Omega^2+V^2/4}T=2\pi$. One only needs to adjust the relative phases of Rydberg lasers at proper times and keep the detuning and the Rabi frequencies constant, which is suitable for experimental realization. It should also be noted that the scheme works effectively with weak Rydberg interaction which allows a long separation distance between two atoms.

The dynamics of the computational basis under the control of pulse sequences (4) is shown in Fig. 2(a).
The population of the basis state $|k\rangle$ is labeled as $P_{k}$. The dynamics of state $|k\rangle$ is obtained by solving Schr\"{o}dinger's Equation with Hamiltonian (1). The parameters are set to $V=2\pi, \kappa=1.65$. As can be seen from the numerical results, for the system prepared in the initial state $|k\rangle$, it will return to the initial population. The accumulated phases $\varphi_k$ of $|k\rangle$ should also be considered, which is obtained by $\varphi_k=angle(\langle \psi_\mathrm{f}|\psi_\mathrm{i}\rangle), |\psi_\mathrm{i}\rangle=|k\rangle$ (the function $angle$ returns the phase of a complex number). As shown in Fig. 2(b), the controlled-phase $\delta\gamma$ can be tilted by $\kappa$, which is related to the ratio between $\Omega$ and $V$. The other parameters are the same as in Fig. 2(a). The $\mathrm{C_z}$ gate can be achieved when $\kappa=1.65$ as marked by a red diamond in Fig. 2(b). It can be found that $\delta\gamma$ turns to be trivial (tend to $-2\pi$) when $\kappa\gg 1$ as the two atoms can be treated individually when $\Omega\gg V$. Therefore, controlled-phase gates in atomic arrays can be realized without Rydberg blockade, and the controlled-phase can be tilted by suitable parameters.

\begin{figure}[ptb]
\begin{center}
\includegraphics[width=8.5cm]{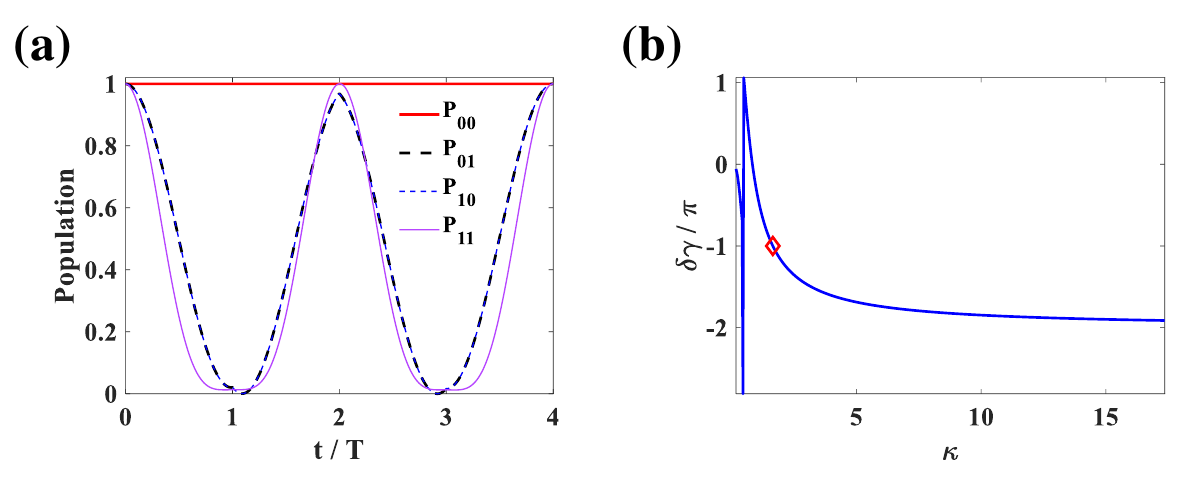}
\caption{\label{fig2}
(a) Population dynamics of $P_k$ under the geometric control with Eq. (4), which is solved by the Schr\"{o}dinger's Equation with Hamiltonian (1). Red-solid line: $P_{00}$, black-dashed line: $P_{01}$, thin blue-dashed line: $P_{10}$, thin purple-solid line: $P_{11}$. As can be seen that all states $|k\rangle$ evolve cyclically. (b) The controlled-phase $\delta\gamma$ versus the variation of $\kappa=\Omega/V$. The $C_\mathrm{z}$ gate can be achieved when $\kappa=1.65$, as labeled by the red diamond.}
\end{center}
\end{figure}

\begin{figure}[ptb]
\begin{center}
\includegraphics[width=8.5cm]{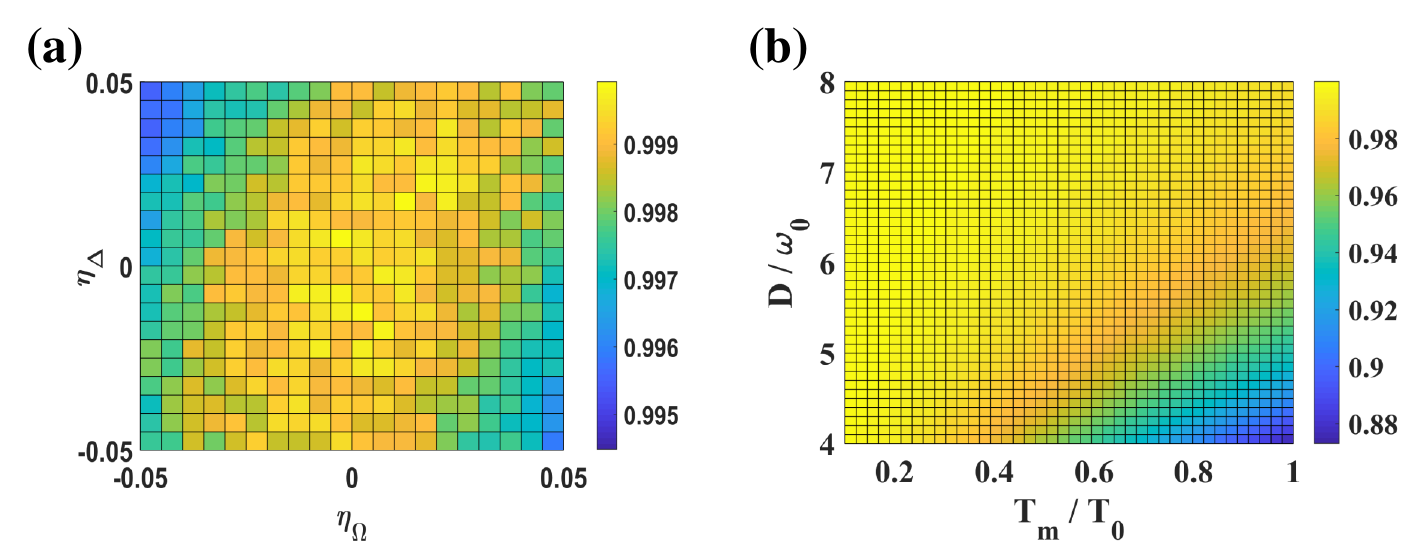}
\caption{\label{fig3}
Robustness of the proposed $\mathrm{C_z}$ gate. (a) Fidelity $F$ versus the variation of Rabi frequencies and detuning by introducing $\Omega'=(1+\eta_\Omega R(t))\Omega$, $\Delta'=(1+\eta_\Delta R(t))\Delta$. $R(t)$ are the random numbers sequences with mean vanishing values. (b) Fidelity $F$ versus the spacing $D$ between the atoms and the atomic temperature $T_m$.}
\end{center}
\end{figure}

\textit{Robustness against random noises and thermal motion.}--- In Fig. 3(a), the robustness against random noises of the proposed scheme are discussed. Without losing generality, we adopt the pulse sequences (4). $\kappa=1.65$ confirms that the $\mathrm{C_z}$ gate can be realized with weak Rydberg interaction. The deviation of the parameters are introduced by $\Omega'=(1+\eta_\Omega R(t))\Omega$, $\Delta'=(1+\eta_\Delta R(t))\Delta$. $R(t)$ is the random numbers sequences with mean vanishing values and modulated by $\eta_{\Omega,\Delta} \in [-0.05, 0.05]$. The fidelity is defined as $F=|tr({U_a}{U_\mathrm{z}}^{\dagger})|/16$, $U_{\mathrm{z}}$ is the ideal matrix of $\mathrm{C_Z}$ gate while ${U_a}$ is the actual evolution operator which is calculated by solving the Schr\"{o}dinger's Equation with Hamiltonian (1). $tr$ symbols the trace of the matrix. As shown in Fig. 3(a), the fidelity $F$ is larger than 0.995 within the fluctuation $5\%$, which is an achievable precision in experiments. Note that the numerical results have been averaged 100 times for randomized evolution. Therefore, the proposed gate is robust to random noise from pulses.

Here we also concern with the robustness of the proposed gate against the thermal motion of the atoms. It will affect the Rydberg interaction strength $V$ due to its distance dependence. The distance between two vibratory atoms can be described as $D=L+br_0\sin(\omega t)$, $L$ is the distance between the equilibrium positions of two atoms. $r_0$ is the waist radius of the optical dipole trap and is set to $1 \mu m$. $\omega$ is the vibration frequency as determined by the geometry of the optical dipole trap. We set $\omega=2\pi/T\times50$ according to typical experimental parameters \cite{Liu12021}. $br_0$ characterizes the synthetic vibrating amplitude of the two atoms that is related to the temperature $T_m$ of the atoms. $b=\sqrt{2T_m/T_{0}}$, $T_0=20 \mu$K corresponds to a typical temperature of a magneto-optical trap \cite{Thompson2013}. Consequently, the Rydberg interacting strength varies with time and scales as $V_a=V(D/L)^6$, after considering the thermal motion of the atoms. By replacing $V$ with $V_a$ in Hamiltonian (1) and adopting the same pulse sequences in Fig. 2(a), fidelity $F$ of $C_\mathrm{z}$ gate that contains thermal motion effects is numerically obtained. $T_m\in[1\mu \mathrm{K}, 20\mu \mathrm{K}]$, $D/r_0\in[4, 8]$ are tested in Fig. 3(b). As can be seen, $F$ will drop quickly as $T_m$ increases when $D/r_0<6$. Meanwhile, $F$ will be higher than 0.99 when $D/r_0>8$, even for $T_m>20 \mu$K (it can be easily obtained after the polarization gradient cooling in typical experiments). As a result, the thermal motion of the atoms will be irrelevant as long as the spacing between the atoms is sufficiently large. It will significantly reduce the crosstalk between the atoms.

\textit{Mechanism of geometric quantum control.}--- As demonstrated in Fig. 3(a), the proposed two-qubit gate is robust against random noises upon the control parameters. In the following, we explain the basic physical mechanics. In fact, the robustness of the quantum gate comes from geometric control of cyclic evolution. The new feature is that the proposed gate consolidates two different kinds of geometric control since the quantum state evolves in a high-dimensional Hilbert space.

The subspace Hamiltonian $H_{11}$ in Eq.2(c) is a three-level $\Lambda$ system with $\Delta=-V/2$. $H_{11}$ can be rewritten as $H_{11}=(\Omega e^{i\varphi}|R\rangle\langle b|+\mathrm{H.c.})+\Delta|R\rangle\langle R|$, where $|b\rangle=\sin(\theta_{11}/2)e^{-2i\varphi}|11\rangle+\cos(\theta_{11}/2)|rr\rangle$ with $\theta_{11}=\pi/2$. An orthogonal state $|d\rangle=\cos(\theta_{11}/2)e^{-2i\varphi}|11\rangle-\sin(\theta_{11}/2)|rr\rangle$ can be introduced for the system. Thus, the diabatic dynamics of $H_{11}$ can be treated as an oscillation between $|R\rangle$ and $|b\rangle$ with Rabi frequency $2\Omega$ and detuning $\Delta$. The system of the subspace will return to the initial population with period $T_{11}=2\pi/\Omega_{\mathrm{11}} (\Omega_{\mathrm{11}}=\sqrt{4\Omega^2+\Delta^2})$ \cite{Sjoqvist2012,Feng2013,Abdumalikov2013}. It can be checked that both $|b\rangle$ and $|d\rangle$ are cyclic states under cyclic driving since $U_{11}|d\rangle=|d\rangle, U_{11}|b\rangle=e^{-i\chi}|b\rangle$ with $U_{11}=U_{11}(\varphi)=e^{-iH_{11}T}$. $\chi$ is the accumulation phase of $|b\rangle$ and it can be verified that $\chi=\frac{\pi\Delta}{\Omega_{\mathrm{11}}}=\frac{\pi}{\sqrt{16\kappa^2+1}}$ \cite{Sjoqvist2016}. It can also be checked that $\langle d|H_{11}|d\rangle=\langle b|H_{11}|b\rangle=0$ which results in vanishing dynamical phases. As a consequence, given an input state $|\psi_{11}\rangle=h_{11}|11\rangle+h_{rr}|rr\rangle=h_d|d\rangle+h_b|b\rangle$, it will experience the non-adiabatic holonomic control $U_{11}$. It should be noted that the geometric phase on the dressed state $|b\rangle$ corresponds to a non-Abelian transformation on the bare states $|11\rangle$ and $|rr\rangle$ as given by
\begin{equation}
    U_{11}(\varphi)=\frac{1}{2}\left(\begin{array}{cc}
        1+e^{-i\chi} & e^{-2i\varphi}(1-e^{-i\chi}) \\
        e^{-2i\varphi}(1-e^{-i\chi}) & 1+e^{-i\chi}  \\
    \end{array}\right).\\
\end{equation}
As a result, for the system prepared in state $|11\rangle$, it will not return to the initial population after the control. To realize the controlled-phase gate, pulse sequences should be adopted, that is, $U'_{11}=U_{11}(\pi/2)U_{11}(0)=e^{-i\chi}I$ will return the system to the initial population and induce a $\chi$ phase shift in state $|11\rangle$ ($I$ is the identity matrix). Therefore, non-adiabaitc holonomic control can be performed in the subspace related to $|11\rangle$.

We also concern the dynamics of $H_{10} (H_{01})$ in Eq.(2b). $\Omega$ is set to constant and $\Delta=-V/2$ to be consistent with the control in $H_{11}$. The evolution operator will be given by
\begin{equation}
    U_{10}(\varphi)=e^{-iH_{10}T}=\cos\beta I-i\sin\beta \mathbf{n}\cdot\mathbf{\sigma}
\end{equation}
where $\mathbf{n}=(\sin\theta_{10}\sin\varphi, \sin\theta_{10}\cos\varphi, \cos\theta_{10})$, $\tan\theta_{10}=\Omega/\Delta$, $\beta=\sqrt{\Omega^2+\Delta^2}T/2$. Similar results can be obtained for the evolution operator $U_{01}$ subjected to $H_{01}$. It can be checked that the cyclic evolution of $|10(01)\rangle$ can be met when $T_{10}=2\pi/\sqrt{\Omega^2+\Delta^2}$. The state $|10(01)\rangle$ will gain a $\pi$ phase shift after evolution. However, one can find $T_{10}\neq T_{11}$ which indicates that the cyclic evolution condition for $H_{10}$ and $H_{11}$ cannot be satisfied simultaneously. Thus, the control task for $H_{10}$ can be defined as: Driving state $|10(01)\rangle$ cyclically with $N$ times of period $T_{11}$ \cite{note}. An effective way to solve the problem is to use composite pulse sequences. For example, using a four-pulse sequence $U'_{10}=U_{10}(\pi/2)U_{10}(0)U_{10}(\pi/2)U_{10}(0)$. An exact solution of the cyclic evolution governed by $U'_{10}$ is found to be $\tan\beta\cos\theta_{10}+1=0$ which results in $\kappa=0.335$. Actually, when $\kappa\gg1$, $\beta, \theta_{10}\rightarrow\pi/2$, $U'_{10}=-I$ which induce a $\pi$ phase shift in state $|10\rangle$. For general cases, the accumulated phase $\phi_{10}$ of $|10\rangle$ is obtained by  $\phi_{10}=angle(\langle\psi_\mathrm{f}^{(2)}|10\rangle)$, $|\psi_\mathrm{f}^{(2)}\rangle$ is the final state after the evolution $U'_{10}$. It can be checked that $\phi_{10}$ is only related to the geometric parameter $\kappa=-\tan\theta_{10}/2$. Since the evolution is cyclic and the total phase $\phi_{10}$ is independent of the dynamical parameters ($\Omega, \Delta, V$), an unconventional geometric control is realized upon state $|10\rangle$ \cite{zhusl2003}. Similar discussions can be also applied to state $|01\rangle$.

\begin{figure}[ptb]
\begin{center}
\includegraphics[width=8.5cm]{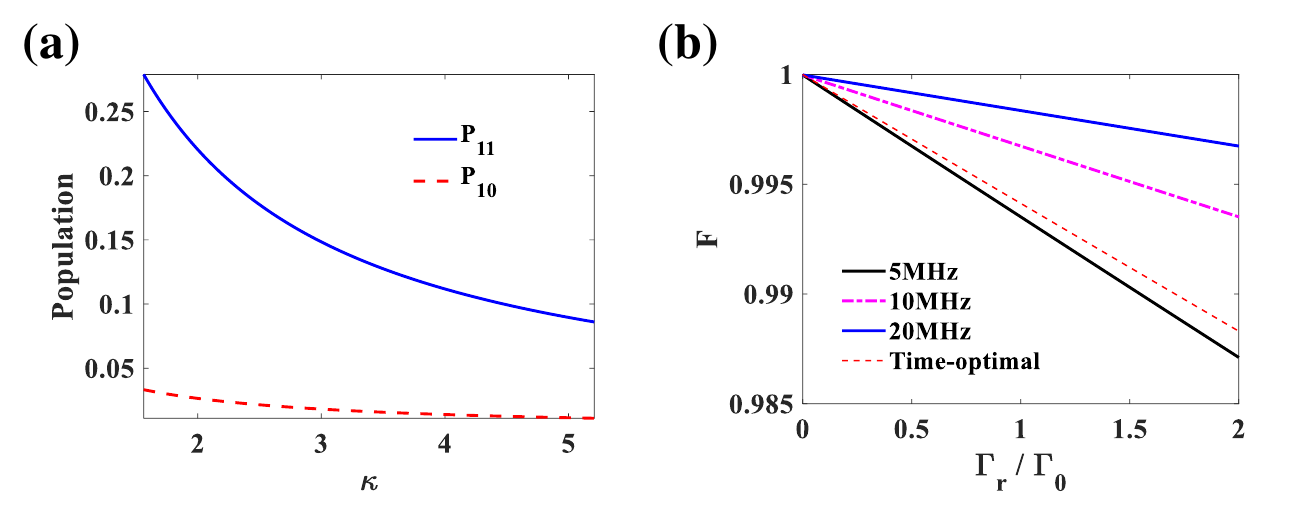}
\caption{\label{fig4}
(a) Interferometric results between the non-adiabatic holonomic phase and the unconventional geometric phase. The interferometry is constructed by Eq.(7) and the relative phases can be tilt by $\kappa$. Blue-solid line: population $P_{11}$, red-dashed line: population $P_{10}$. (b) Fidelity $F$ of $C_\mathrm{z}$ gate against decay rate $\Gamma_r$ of the Rydberg states. The scheme performs as good as the time-optimal scheme with the same Rabi frequency $\Omega$ and can be improved by increasing $\Omega$. Black-solid line, pink dashed-dotted line, blue solid line: $F$ of the proposed gate with $\Omega=2\pi\times5, 10, 20$ MHz. Red-dashed line: $F$ of the time-optimal gate with $\Omega=2\pi\times5$ MHz.  }
\end{center}
\end{figure}

\textit{Unconventional geometric interferometry.}--- In Fig. 4(a), we discuss the realization of a new type of interference where two different kinds of geometric phases can interfere \cite{Falci2000,Du2003,Peng2010,Chen2019,Yang2019}. Actually, utilizing the high dimensions of the interacting model, non-adiabatic holonomic phase (acquired in a three-level system) and the unconventional geometric phase (acquired in a two-level system) can interfere in the discussed model. Given an initial state $|10\rangle$, the pulses of constructing the interferometer are described as
\begin{equation}
    B_{\mathrm{geo}}=B_{\pi/2}B_{\mathrm{inter}}B_{\pi/2},
\end{equation}
where the first $B_{\pi/2}$ pulses to prepare the system to the equally weighted superposition states between $|10\rangle$ and $|11\rangle$. $B_{\pi/2}=(|10\rangle\langle11|+|11\rangle\langle10|)/\sqrt{2}=I_1\otimes Q_{\pi/2}$, $Q_{\pi/2}$ is the pulse to prepare atom 2 in an equally weighted superposition state between $|0\rangle_2$ and $|1\rangle_2$. After that, pulse $B_{\mathrm{inter}}$ drive states $|10\rangle$ and $|11\rangle$ evolve inside the subspace governed by $H_{10}$ and $H_{11}$, respectively. The last $B_{\pi/2}$
pulse is used to induce interference within the interferometer. To realize geometric control upon $|10\rangle$ and $|11\rangle$, $B_{\mathrm{inter}}$ adopts the waveforms in Eq. (4) in time $t: 0\rightarrow T$ with the control parameters as in Fig. 2(a). When $\kappa>1$, $T_{10}\approx T_{11}$. The evolution of $|10\rangle$ and $|11\rangle$ satisfies the cyclic condition simultaneously. As in Fig. 4(a), the population after evolution versus $\kappa$ is plotted, red-dashed line: population of $|10\rangle$, $P_{10}$; blue-solid line: population of $|11\rangle$, $P_{11}$. By tilting $\kappa$, the accumulated phase of $|10(11)\rangle$ will change and can be detected by the interferometer. Due to the non-Abelian character of Eq. (5), the population $P_{10}+P_{11}\neq 1$, which is different from the Abelian interferometer \cite{Falci2000,Du2003,Peng2010}. Therefore, the high-dimensional Hilbert space allows us to investigate new types of interferometry.

Based on the above geometric control, Eq.(4) that used to realize the controlled-phase gate is designed. In Fig. 4(b), fidelity $F$ of $C_\mathrm{z}$ gate against the decoherence effect is discussed. Here we adopt the experimentally feasible parameters as: $\Omega=2\pi\times m$ MHz ($m=5, 10, 20$), $V=\Omega/1.65$. We consider Rydberg states with finite spontaneous emission rate $\Gamma_r$ and the decoherence effect is introduced by replacing $\Delta$ by $\Delta'=\Delta+i\Gamma_r$ in Hamiltonian (1), $\Gamma_r=r\Gamma_0$, $\Gamma_0=2\pi\times10$ kHz. $F$ is derived by $F=|tr(\rho_\mathrm{f}\rho_\mathrm{i})|$, $\rho_\mathrm{i}=|\psi_\mathrm{i}\rangle\langle\psi_\mathrm{i}|$,  $|\psi_\mathrm{i}\rangle=\sum_k|k\rangle/2$ is the initial state. $\rho_\mathrm{f}$ is the final density matrix as solved by the density matrix equation with Hamiltonian (1). As shown in Fig. 4(b), $F$ decreases as the decay rate increases. The robustness against Rydberg states decay can be enhanced when the Rabi frequency $\Omega$ is increased as the total evolution time is shorten; blue-solid line: $m = 5$ MHz,  purple dashed-dotted line: $m = 10$ MHz, black-solid line: $m = 20$ MHz. We also make a comparison of controlled-phase gate realized by geometric control and the time-optimal method (working at the Rydberg blockade region). The waveforms of the time-optimal scheme are chosen to be:
\begin{equation}
    \Omega_t=2\pi\times5 \mathrm{MHz}, \varphi_t=A\cos(\omega t-\varphi_0),
\end{equation}
where $A=2\pi\times0.1122$, $\omega=1.4031\Omega_t$, $\varphi_0=-0.7318$. The evolution time of time-optimal scheme is given by $T_t=2.43\pi/\Omega_t$. The Rydberg interaction strength of the time-optimal scheme is set to $V_t=2\pi\times450$ MHz which can be realized by placing two atoms with distance of $L_t=2 \mu$m and $n_r=53$, according to Ref.\cite{Evered2023}.  The calculation of $F$ of the time-optimal scheme is the same as the geometric control scheme.  Due to the non-adiabatic character of the geometric control ($4T\sim T_t$), the geometric control against decay of performs as well as the time-optimal scheme under the same Rabi frequency. When the Rydberg interaction strength is $V=2\pi\times3$ MHz, the spacing between two atoms of the geometric control is increased to 4.6 $\mu$m ($L^6$ scaling law) with $n_r=53$, which will strongly reduce the crosstalk between the atoms.

\begin{figure}[ptb]
\begin{center}
\includegraphics[width=4.5cm]{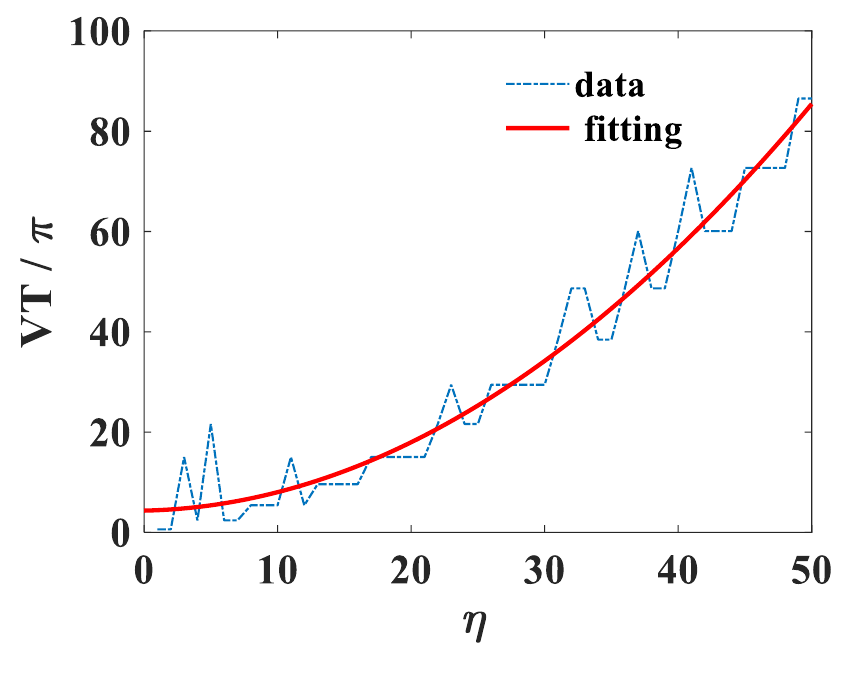}
\caption{\label{fig5}
The actuating quantity ($VT$) to realize high fidelity $C_\mathrm{z}$ gate against interaction strength $V$. The increase of the interaction strength $V$ will induce a quadratic growth in the actuating quantity.}
\end{center}
\end{figure}

\textit{Minimal interaction energy for the two-qubit gates.} --- Finally, we want to discuss the relationship between the actuating quantity $VT$ used to realize the $C_\mathrm{z}$ gate and the interaction strength $V$ \cite{Vidal2002,Carlini2007,Ashhab2012}. A smaller $VT$ indicates that the $C_\mathrm{z}$ gate can be implemented with less physical resources. The control waveforms follow Eq.(4) where the interaction strength is varied by $V=\eta V_0$, $V_0=2\pi$. The definition of $F$ is the same that in Fig. 3. For specific value of $V$, we adjust the phase $\varphi$ and $T$ to find the region fulfill $F>0.96$. If several $(\varphi, T)$ fulfill the requirement of $F>0.96$, we average $T$ to give the effective actuating quantity $VT$. Blue dashed-dotted line: numerical solution of realizing $C_\mathrm{z}$ gate; Red-solid line: fitting curve (quadratic function ) for the data. As can be seen, the proposed scheme works effectively in a wide range of $V$. Furthermore, results of Fig. 5 reveals that increasing the interaction strength $V$ will induce a quadratic growth in the actuating quantity. The above result can be understood by recognizing that the transition probability amplitude of the Rydberg state is proportional to $\Omega/V$. To accumulate enough phase from the interaction, the evolution time $T$ should be proportional to $V$. As a result, $VT$ increases quadratically as $V$ increases. In our proposal, the condition for realizing the $C_\mathrm{z}$ gate is $4VT=2.4\pi$, which is dozens of times smaller than the scheme that work in the Rydberg blockade region \cite{Evered2023}. Therefore, the using of weak Rydberg interaction in the controlled-phase gate will significantly reduce the physical resources.


\textit{Conclusion.}--- In summary, we have proposed a scheme for realizing a two-qubit controlled-phase gate in an atomic array without Rydberg blockade. The proposal can work with large atomic spacing and low lying Rydberg states. It will significantly reduce the crosstalk between the atoms and the dephasing effect (mainly due to the finite $T_2$ which is proportional to $1/n_r^7$ \cite{Evered2023}). The scheme is robust against random noises due to the geometric characteristic and operates as fast as the time-optimal scheme based on the non-adiabatic evolution. Since the proposed controlled-phase gate consolidates the non-adiabatic holonomic control and the unconventional geometric control simultaneously, it provides a platform to investigate the interference between two different types of geometric phases. Therefore, our proposal provides a fast and robust way to realize quantum computation in an atomic array with Rydberg interactions. Furthermore, our work may also trigger the investigation of new-type geometric gates in high-dimension Hilbert space (i.e., adiabatic geometric/adiabatic non-Abelian-mixed multi-qubit gates).

\textit{Acknowledgment.} --- This work was supported by the National Natural Science Foundation of China (Grants No. 12074132, and No. U20A2074), NSF of Guangdong province (Grant No. 2024A1515012516), Guangdong Provincial Quantum Science Strategic Initiative (Grant No. GDZX2303006). Y. Ming and Z. X. Fu contribute equally to this work.


\begin{thebibliography}{99}

\bibitem{Saffman2016}M. Saffman, Quantum computing with atomic qubits and Rydberg interactions: Progress and challenges, J. Phys. B: At. Mol. Opt. Phys. {\bf 49}, 202001 (2016).
\bibitem{Henriet2020}L. Henriet, L. Beguin, A. Signoles, T. Lahaye, A. Browaeys, G. O. Reymond, and C. Jurczak, Quantum computing with neutral atoms, Quantum {\bf 4}, 327 (2020).
\bibitem{Browaeys2020}A. Browaeys and T. Lahaye, Many-body physics with individually controlled Rydberg atoms, Nat. Phys. {\bf 16}, 132 (2020).
\bibitem{Scholl2021}P. Scholl, M. Schuler, H. J. Williams, A. A. Eberharter, D. Barredo, K.-N. Schymik, V. Lienhard, L.-P. Henry, T. C. Lang, T. Lahaye, A. M. L\"{a}uchli, and A. Browaeys, Quantum simulation of 2D antiferromagnets with hundreds of Rydberg atoms, Nature (London) {\bf 595}, 233 (2021).
\bibitem{Ebadi2021}S. Ebadi, T. T. Wang, H. Levine, A. Keesling, G. Semeghini, A. Omran, D. Bluvstein, R. Samajdar, H. Pichler, W. W. Ho, S. Choi, S. Sachdev, M. Greiner, V. Vuleti, and M. D. Lukin, Quantum phases of matter on a 256-atom programmable quantum simulator, Nature (London) {\bf 595}, 227 (2021).
\bibitem{Kim2024}K. H. Kim, F. Yang, K. M{\o}lmer, and J. W. Ahn, Realization of an extremely anisotropic heisenberg magnet in Rydberg atom arrays, Phys. Rev. X {\bf 14}, 011025 (2024).


\bibitem{Levine2019}H. Levine, A. Keesling, G. Semeghini, A. Omran, T. T. Wang, S. Ebadi, H. Bernien, M. Greiner, V. Vuletic, H. Pichler, and M. D. Lukin, Parallel implementation of high-fidelity multiqubit gates with neutral atoms, Phys. Rev. Lett. {\bf123}, 170503 (2019).
\bibitem{Madjarov2020}I. S. Madjarov, J. P. Covey, A. L. Shaw, J. Choi, A. Kale, A. Cooper, H. Pichler, V. Schkolnik, J. R. Williams, and M. Endres, High-fidelity entanglement and detection of alkaline-earth Rydberg atoms, Nat. Phys. {\bf16}, 857
(2020).
\bibitem{Fu2022}Z. Fu, P. Xu, Y. Sun, Y. Y. Liu, X. D. He, X. Li, Min Liu, R. B. Li, J. Wang, L. Liu, and M. S. Zhan, High-fidelity entanglement of neutral atoms via a Rydberg-mediated single-modulated-pulse controlled-phase gate, Phys. Rev. A {\bf105}, 042430 (2022).
\bibitem{Evered2023}S. J. Evered, D. Bluvstein, M. Kalinowski, S. Ebadi, T. Manovitz, H. Y. Zhou, S. H. Li, A. A. Geim, T. T. Wang, N. Maskara, H. Levine, G. Semeghini, M. Greiner, V. Vuletic, and M. D. Lukin, High-fidelity parallel entangling gates on a neutral atom quantum computer, Nature (London) {\bf 622}, 268 (2023).
\bibitem{Bluvstein2024}D. Bluvstein, et al., Logical quantum processor based on reconfigurable atom arrays, Nature (London) {\bf 626}, 58 (2024).

\bibitem{Saffman2010}M. Saffman, T. G. Walker, and K. M{\o}lmer, Quantum information with Rydberg atoms, Rev. Mod. Phys. {\bf82}, 2313 (2010).
\bibitem{Jaksch2000}D. Jaksch, J. I. Cirac, P. Zoller, S. L. Rolston, R. C\^{o}t\'{e}, and M. D. Lukin, Fast quantum gates for neutral atoms, Phys. Rev. Lett. {\bf85}, 2208 (2000).
\bibitem{Isenhower2010}L. Isenhower, E. Urban, X. L. Zhang, A. T. Gill, T. Henage, T. A. Johnson, T. G. Walker, and M. Saffman, Demonstration of a neutral atom Controlled-NOT quantum gate, Phys. Rev. Lett. {\bf104}, 010503 (2010).

\bibitem{Jau2016}Y. Y. Jau, A. M. Hankin, T. Keating, I. H. Deutsch, and G. W. Biedermann, Entangling atomic spins with a Rydberg-dressed spin-flip blockade, Nat. Phys. {\bf12}, 71 (2016).
\bibitem{Shi2018}X. F. Shi, Deutsch, Toffoli, and CNOT gates via Rydberg blockade of neutral atoms, Phys. Rev. Appl. {\bf9}, 051001 (2018).
\bibitem{Liu2020}B. J. Liu, S. L. Su, and M. H. Yung, Nonadiabatic noncyclic geometric quantum computation in Rydberg atoms, Phys. Rev. Res. {\bf2}, 043130 (2020).
\bibitem{Theis2016}L. S. Theis, F. Motzoi, and F. K. Wilhelm, High-fidelity Rydberg-blockade entangling gate using shaped, analytic pulses, Phys. Rev. A {\bf94}, 032306 (2016).
\bibitem{Wilk2010}T. Wilk, A. Ga\"{e}tan, C. Evellin, J. Wolters, Y. Miroshnychenko, P. Grangier, and A. Browaeys, Entanglement of two individual neutral atoms using Rydberg blockade, Phys. Rev. Lett. {\bf104}, 010502 (2010).
\bibitem{Sun2020}Y. Sun, P. Xu, P. X. Chen, and L. Liu, Controlled-phase gate protocol for neutral atoms via off-resonant modulated driving, Phys. Rev. Res. {\bf13}, 024059 (2020).
\bibitem{Zeng2017}Y. Zeng, P. Xu, X. He, Y. Liu, M. Liu, J. Wang, D. J. Papoular, G. V. Shlyapnikov, and M. Zhan, Entangling two individual atoms of different isotopes via Rydberg blockade, Phys. Rev. Lett. {\bf119}, 160502 (2017).
\bibitem{Li2022}X. X. Li, X. Q. Shao, and W. B. Li, Single temporal-pulse modulated parameterized controlled-phase gate for Rydberg atoms, Phys. Rev. Appl. {\bf18}, 044042 (2022).

\bibitem{Sjoqvist2008}E. Sj\"{o}qvist, A new phase in quantum computation, Physics {\bf 35}, 1 (2008).
\bibitem{Berry1984}M. V. Berry, Quantal phase factors accompanying adiabatic changes, Proc. R. Soc. London A {\bf 392}, 45 (1984).
\bibitem{Moller2008}D. M{\o}ller, L. B. Madsen, and K. M{\o}lmer,  Quantum gates and multiparticle entanglement by
Rydberg excitation blockade and adiabatic passage, Phys. Rev. Lett. {\bf100}, 170504 (2008).
\bibitem{wilzeck1984}F. Wilczek and A. Zee, Appearance of gauge structure in simple dynamical systems, Phys. Rev. Lett. {\bf 52}, 2111 (1984).
\bibitem{duan2001}L. M. Duan, J. I. Cirac, and P. Zoller, Geometric manipulation of trapped ions for quantum computation, Science, {\bf292}, 1695 (2001).
\bibitem{Aharonov1987}Y. Aharonov and J. Anandan, Phase change during a cyclic quantum evolution, Phys. Rev. Lett. {\bf 58}, 1593 (1987).
\bibitem{SLZhu2002}S. L. Zhu and Z. D. Wang, Implementation of universal Quantum gates based on nonadiabatic geometric phases, Phys. Rev. Lett. {\bf 89}, 097902 (2002).
\bibitem{zhusl2003}S. L. Zhu and Z. D. Wang, Unconventional geometric quantum computation, Phys. Rev. Lett. {\bf 91}, 187902 (2003).
\bibitem{Sjoqvist2012}E. Sj\"{o}qvist, D. M. Tong, L. M. Andersson, B. Hessmo, M. Johansson, and K. Singh, Non-adiabatic holonomic quantum computation, New J. Phys. {\bf14}, 103035 (2012).
\bibitem{Feng2013}G. Feng, G. Xu, and G. Long, Experimental realization of nonadiabatic holonomic quantum computation, Phys. Rev. Lett. {\bf110}, 190501 (2013).
\bibitem{Abdumalikov2013}A. A. Abdumalikov Jr, J. M. Fink, K. Juliusson, M. Pechal, S. Berger, A. Wallraff, and S. Filipp, Experimental realization of non-Abelian non-adiabatic geometric gates, Nature (London) {\bf496}, 482 (2013).
\bibitem{Liang2014}Z. T. Liang, Y. X. Du, W. Huang, Z. Y. Xue, and H. Yan, Nonadiabatic holonomic quantum computation in decoherence-free subspaces with trapped ions, Phys. Rev. A {\bf 89}, 062312 (2014).
\bibitem{Sjoqvist2016}E. Sj\"{o}qvist, Nonadiabatic holonomic single-qubit gates in off-resonant $\Lambda$ systems, Phys. Lett. A {\bf 380}, 65 (2016).
\bibitem{Ma2023}Z. Ma, J. W. Xu, T. Chen, Y. Zhang, W. Zheng, S. X. Li, D. Lan, Z. Y. Xue, X. S. Tan, and Y. Yu, Noncyclic nonadiabatic geometric quantum gates in a superconducting circuit, Phys. Rev. Appl. {\bf 20}, 054047 (2023).

\bibitem{Chen2024}Z. Y. Chen, J. H. Liang, Z. X. Fu, H. Z. Liu, Z. R. He, M. Wang, Z. W. Han, J. Y. Huang, Q. X. Lv, and Y. X. Du, Single-pulse two-qubit gates for Rydberg atoms with noncyclic geometric control, Phys. Rev. A {\bf109}, 042621 (2024).
\bibitem{Sun2023}Y. Sun, Off-resonant modulated driving gate protocols for two-photon ground-Rydberg transition and finite Rydberg blockade strength, Opt. Exp. {\bf31}, 3114, (2023).

\bibitem{Beterov2016} I. I. Beterov, M. Saffman, E. A. Yakshina, D. B. Tretyakov, V. M. Entin, S. Bergamini, E. A. Kuznetsova, and I. I. Ryabtsev, Two-qubit gates using adiabatic passage of the Stark-tuned Forster resonances in Rydberg atoms, Phys. Rev. A {\bf94}, 062307 (2016).
\bibitem{Pelegri2022}G. Pelegr, A. J. Daley, and J. D. Pritchard, High-fidelity multiqubit Rydberg gates via two-photon adiabatic rapid passage, Quantum Sci. Technol. {\bf7}, 045020 (2022).
\bibitem{Mitra2020}A. Mitra, M. J. Martin, G.W. Biedermann, A. M. Marino, P. M. Poggi, and I. H. Deutsch, Robust M{\o}lmer-S{\o}rensen gate for neutral atoms using rapid adiabatic Rydberg dressing, Phys. Rev. A {\bf101}, 030301(R) (2020).




\bibitem{Kang2018}Y. H. Kang, Y. H. Chen, Z. C. Shi, B. H. Huang, J. Song, and Y. Xia, Nonadiabatic holonomic quantum computation using Rydberg blockade, Phys. Rev. A {\bf97}, 042336 (2018).
\bibitem{Jandura2022}S. Jandura and G. Pupillo, Time-optimal two- and three-qubit gates for Rydberg atoms, Quantum {\bf6}, 712 (2022).

\bibitem{Liu12021}Y. Y. Liu, Y. Sun, Z. Fu, P. Xu, X. Wang, X. D. He, J. Wang, and M. S. Zhan, Infidelity induced by ground-rydberg decoherence of the control qubit in a two-qubit Rydberg-blockade gate, Phys. Rev. Appl. {\bf15}, 054020 (2021).
\bibitem{Thompson2013}J. D. Thompson, T. G. Tiecke, A. S. Zibrov, V. Vuleti\'{c}, and M. D. Lukin, Coherence and raman
sideband cooling of a single atom in an optical tweezer, Phys. Rev. Lett. {\bf110}, 133001 (2013).

\bibitem{note}Since the cyclic evolution of state $|11\rangle$ is more difficult to be fulfilled (only achieved with $U_{11}(0)U_{11}(\pi/2)$), we set the total evolution time to be $T=NT_{11}$.

\bibitem{Falci2000}G. Falci, R. Fazio, G. M. Palma, J. Siewert, and V. Vedral, Detection of geometric phases in superconducting nanocircuits, Nature (London) {\bf407}, 355 (2000).
\bibitem{Du2003}J. F. Du, P. Zou, M. J. Shi, L. C. Kwek, J. W. Pan, C. H. Oh, A. Ekert, D. K. L. Oi, and M. Ericsson, Observation of geometric phases for mixed states using NMR interferometry, Phys. Rev. Lett. {\bf91}, 100403 (2003).
\bibitem{Peng2010}X. H. Peng, S. F. Wu, J. Li, D. Suter, and J. F. Du, Observation of the ground-state geometric phase in a Heisenberg XY model, Phys. Rev. Lett. {\bf105}, 240405 (2010).
\bibitem{Chen2019}Y. T. Chen, R. Y. Zhang, Z. F. Xiong, Z. H. Hang, J. S. Li, J. Q. Shen, and C. T. Chan, Non-Abelian gauge field optics, Nat. Commun. {\bf10}, 3125 (2019).
\bibitem{Yang2019}Y. Yang, C. Peng, D. Zhu, H. Buljan, J. D. Joannopoulos, B. Zhen, and M. Solja\v{c}i\'{c}, Synthesis and observation of non-Abelian gauge fields in real space, Science, {\bf365}, 1021 (2019).

\bibitem{Vidal2002}G. Vidal, K. Hammerer, and J. I. Cirac, Interaction cost of nonlocal gates, Phys. Rev. Lett. {\bf88}, 237902 (2002).
\bibitem{Carlini2007}A. Carlini, A. Hosoya, T. Koike, and Y. Okudaira,  Time-optimal unitary operations, Phys. Rev. A {\bf75}, 042308 (2007).
\bibitem{Ashhab2012}S. Ashhab, P. C. de Groot, and F. Nori, Speed limits for quantum gates in multiqubit systems, Phys. Rev. A {\bf85}, 052327 (2012).





\end{thebibliography}
\end{document}